\documentclass[conference]{IEEEtran}
\IEEEoverridecommandlockouts
\usepackage{cite}
\usepackage{amsmath,amssymb,amsfonts}
\usepackage{algorithmic}
\usepackage{graphicx}
\usepackage{textcomp}
\usepackage{xcolor}
\usepackage{float}
\def\BibTeX{{\rm B\kern-.05em{\sc i\kern-.025em b}\kern-.08em
    T\kern-.1667em\lower.7ex\hbox{E}\kern-.125emX}}
\begin{document}

\title{FPGA-based Toeplitz Strong Extractor for Quantum Random Number Generators\\
\thanks{* Contributed equally to the work \\ The authors acknowledge the support from DIAT(DU) under the grant-in-aid program.}
}

\author{
\IEEEauthorblockN{Shubham Chouhan*}
\IEEEauthorblockA{\textit{School of Quantum Technology},\\ \textit{Defence Institute of Advanced Technology},\\
Pune, India \\
sschouhan625@gmail.com}
\and\and\and\and\and\and\and\and\and\and\and\and\and\and\and\and\and\and\and\and\and\and\and\and\and\and\and\and\and\and\and\and\and\and\and\and\and\and\and\and\and\and\and\and\and\and\and\and\and\and\and\and\and\and\and\and\and\and\and\and\and\and\and\and\and\and\and\and\and\and\and\and\and\and\and\and\and\and\and\and\and\and\and\and\and\and\and\and\and\and\and\and\and\and\and\and\and\and\and\and\and\and\and\and\and\and\and\and\and\and\and\and\and\and\and\and\and\and\and\and\and
\IEEEauthorblockN{Anurag K. S. V.*}
\IEEEauthorblockA{\textit{School of Quantum Technology},\\ \textit{Defence Institute of Advanced Technology},\\
Pune, India \\
anurag.krovvidi@gmail.com}
\and\and\and\and\and\and\and\and\and\and\and\and
\IEEEauthorblockN{G. Raghavan}
\IEEEauthorblockA{\textit{School of Quantum Technology},\\ \textit{Defence Institute of Advanced Technology},\\
Pune, India \\
graghavan@diat.ac.in}
\and\and\and\and\and\and\and\and\and\and\and\and\and
\IEEEauthorblockN{Kanaka Raju P.}
\IEEEauthorblockA{\textit{School of Quantum Technology},\\ \textit{Defence Institute of Advanced Technology},\\
Pune, India \\
raju@diat.ac.in}
}

\maketitle

\begin{abstract}
Quantum Random Number Generators (QRNGs) serve as high-entropy sources for Quantum Key Distribution (QKD) systems. However, the raw data from these quantum sources require post-processing to achieve a nearly uniform distribution. This work presents a state-of-the-art implementation of the Toeplitz Strong Extractor on an FPGA, achieving a benchmark extraction speed of 26.57 Gbps. A detailed implementation flow of the post-processing on the FPGA is provided, along with the execution speeds obtained for different randomness extraction ratios. Raw data from an in-house phase noise-based QRNG is processed on the FPGA using this implementation, and the output is validated using the NIST STS 2.1.2 statistical randomness test suite.
\end{abstract}

\begin{IEEEkeywords}
Toeplitz Hashing, Strong Extractor, Randomness Extraction, QRNG Post Processing 
\end{IEEEkeywords}

\section{Introduction}

Quantum Random Number Generators (QRNGs) play a vital role in cryptographic applications, acting as a major sub-system in Quantum Key Distribution (QKD) Systems. These systems are broadly classified into two types, namely Device Independent-QRNGs (DI-QRNGs) and Device Dependent-QRNGs, more commonly known as Trusted Device-QRNGs (TD-QRNGs). While DI-QRNGs and their various variations offer better security, their random number generation rates are often too low for any real-time applications \cite{cheng_2022_diqrng, cheng_2024_diqrng, li_2024_diqrng, nie_2024_diqrng}. While the development of such DI-QRNGs has steadily progressed over the years, TD-QRNGs have matured faster, resulting in higher random number generation rates, facilitating industry adoption \cite{bai_2021_188, ng_2023_240, marangon_2024_selfiprc, tanizawa_2024_realtime}. Major physical realizations of these systems stem from optics-based phenomena such as single photon sources and beam splitters \cite{jennewein_2000_beamsplitterqrng}, optical parametric oscillators \cite{sunada_2011_opoqrng, marandi_2012_opoqrng}, vacuum fluctuations \cite{huang_2019_homodyneqrng, gehring_2021_homodyneqrng, wang_2023_homodyneqrng, bru_2023_homodyneqrng}, phase noise \cite{nie_2015_the, raffaelli_2018_phasenoiseqrng, roman_2023_phasenoiseqrng}, etc. The raw data generation rates of these systems vary widely \cite{kollmitzer_2020_quantum}. For high-speed random number generation, vacuum fluctuations-based QRNGs \cite{tanizawa_2024_realtime} or Phase Noise-based \cite{nie_2015_the} QRNGs are preferred. Recent developments following optoelectronic integrated circuits and on-chip realizations of these devices have reported generation rates up to 100 Gbps \cite{bru_2023_homodyneqrng} and 240 Gbps \cite{ng_2023_240} respectively.

While the raw data rate of such devices is high, the output from most of these TD-QRNGs require robust post-processing through a randomness extractor to produce theoretically provable randomness \cite{ma_2013_postprocessing}. The standard method to perform post-processing of QRNG raw data follows min-entropy evaluation, which is dependent on the physical system used to realize QRNG \cite{nie_2015_the, bai_2021_188}, followed by randomness extraction based on strong extractors and the leftover hash lemma as described by \cite{ma_2013_postprocessing, kollmitzer_2020_quantum}. One such major extractor is the Toeplitz Strong Extractor (TSE), which has been employed for its ease of use and comparatively lower computational complexity, aiding higher extraction speeds \cite{guo_2024_parallel}. Toeplitz Strong Extractor is primarily based on the binary matrix multiplication operation or hashing between a pseudo-random seed and the raw data acquired from a QRNG \cite{ma_2013_postprocessing, guo_2024_parallel}. Early realizations of the Toeplitz Strong Extractor have resulted in post-processing speeds of 441 Kbps, as shown by \cite{xu_2012_ultrafast}. The algorithmic structure of the Toeplitz Strong Extractor allows for massively parallel binary matrix multiplication operations, due to which a hardware-centric realization of TSE on FPGA was adopted \cite{zhang_2016_fpga}. This resulted in a major speedup in the extraction rates in the order of Gbps as shown by \cite{zhang_2016_fpga, zheng_2019_6, bai_2021_188}. Recent realization of these devices via parallelization of the entropy source, followed by using multiple FPGA boards for randomness extraction, has resulted in a collective post-processing speed of 50 Gbps in multi-channel and 12.5 Gbps in single-channel as reported by \cite{tanizawa_2024_realtime}.

While there has been a steady increase in the efficiency and extraction rates of randomness extraction performed as part of the post-processing of QRNGs, there is a growing need for more efficient implementations of TSE to achieve better extraction speeds to catch up with the raw data generation rates of these devices. This work improves upon the existing implementations of the Toeplitz Strong Extractor by extending it into a high-efficiency FPGA-based post-processing scheme. The details of the methodology are described in Section II, and the results of the randomness extraction and NIST test suite results are shown in Section III. Finally, the concluding remarks and future scope of this work are shared in Section IV. 

\section{Methodology and Implementation Details}

The raw data is collected ($8 \times 10^{5}$ bits) from an in-house phase noise-based QRNG outfitted with an 8-bit ADC to perform the post-processing using the Toeplitz Strong Extractor algorithm on the FPGA. Due to the presence of classical noise and correlations within raw data, it is necessary to perform post-processing on the raw data bits to distill the quantum randomness from it as detailed in \cite{ma_2013_postprocessing, guo_2024_parallel}. First, we evaluate the min-entropy to determine the amount of randomness that can be extracted from raw input data. The length of the output bit-string $(m)$ is then calculated using the leftover hash lemma, the theoretical background of which can be found in \cite{ma_2013_postprocessing, kollmitzer_2020_quantum, guo_2024_parallel}.

\subsection{Seed Generation based on Linear Feedback Shift Register Method}

\begin{figure}[htbp]
\centerline{\includegraphics[width=3.375in]{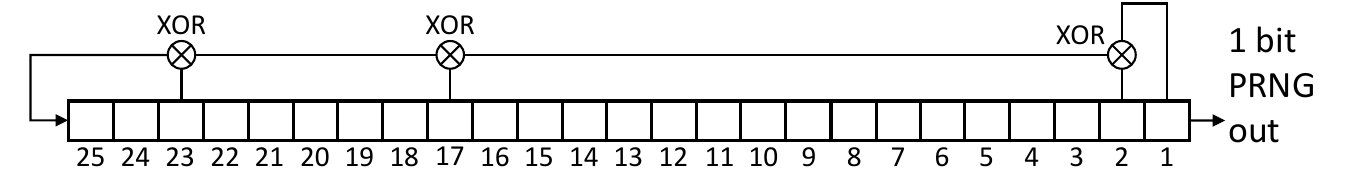}}
\caption{Implementation of LFSR-based PRNG on FPGA.}
\label{fig1}
\end{figure}

Topelitz strong extractors require a random seed to generate the Topelitz string $(tr)$. Here, the Random Seed is obtained via an LFSR-based PRNG implemented on the FPGA. The LFSR-based PRNG is implemented using the following configurations: A 25-bit LFSR seed is chosen, out of which $23$ bits are randomly selected from the sample raw data, and two bits are fixed to binary `1' to avoid all $25$ bits being zero. The LFSR output is taken from the least significant bit, and on each clock cycle, one PRNG bit is generated as shown in Fig.~\ref{fig1}. These bits are concatenated to construct a Toeplitz String $(ts)$ of length $(bs+m-1)$ where $(bs)$ is the length of the block size considered, which is $1 \times 10^{3}$ bits.

\subsection{Toeplitz Strong Extractor}

\begin{figure}[htbp]
\centerline{\includegraphics[width=3.375in]{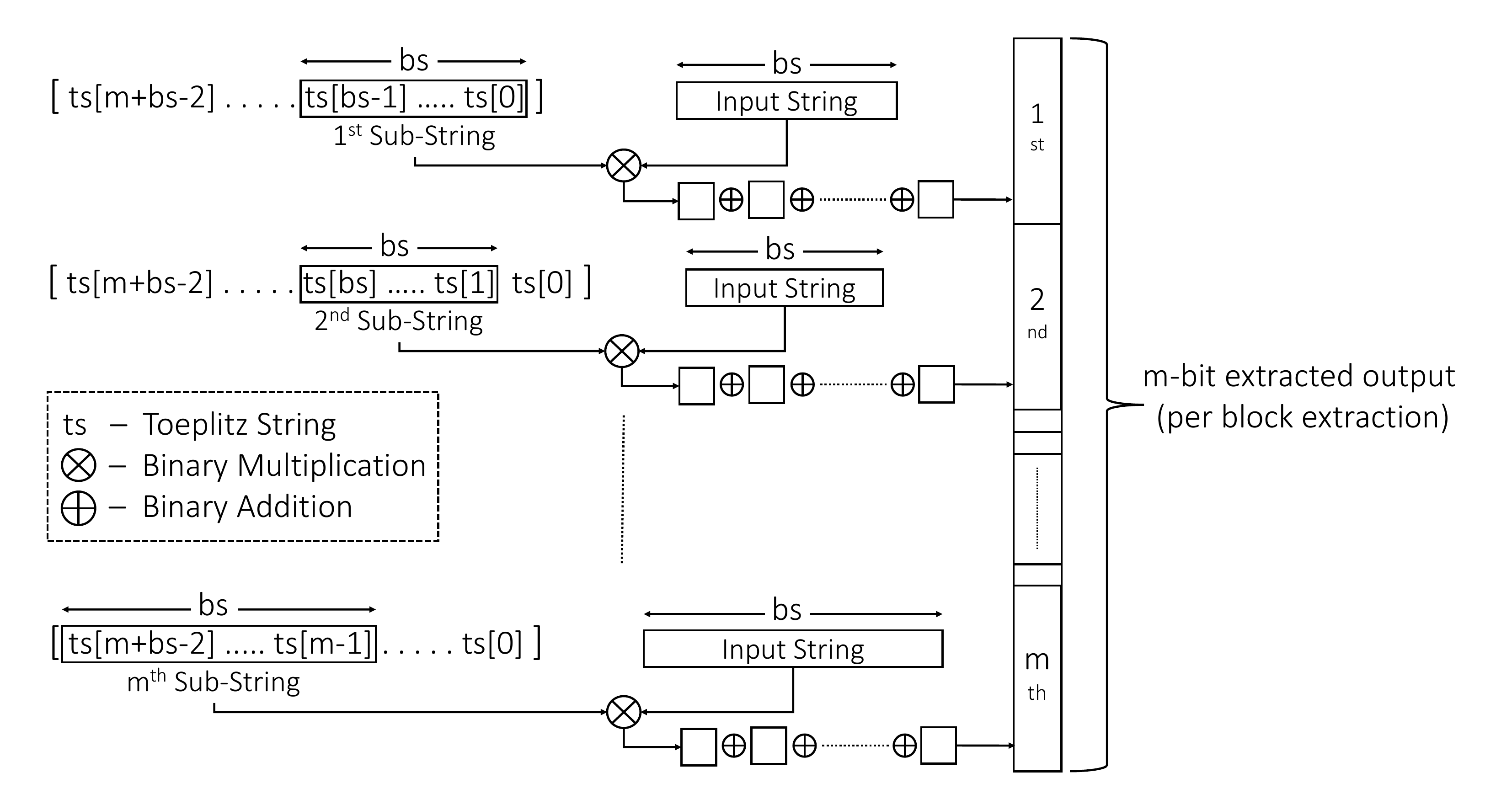}}
\caption{Implementation of Toeplitz Strong Extractor per Block.}
\label{fig2}
\end{figure}

As mentioned earlier, we considered a sample size of $8 \times 10^{5}$ raw QRNG data and block size $(bs)$ of $1 \times 10^{3}$. Due to the massive logical resources available in the FPGA board (Xilinx VC709), we work with $40$ blocks $(K)$ in parallel to perform the randomness extraction via the Toeplitz Strong Extractor. As a result, we are working with a batch size $(L)$ of $4 \times 10^{4}$, and the Toeplitz Strong Extractor is realized per block. To extract bits from input raw data, we need to multiply them with the Toeplitz matrix which can be generated from the Toeplitz String $(ts)$, and due to the unique symmetry (descending diagonal elements from left to right are the same) of the Toeplitz matrix, we can use Toeplitz String $(ts)$ and perform shifting operation per clock cycle to get sub-strings as shown in Fig.~\ref{fig2}, these sub-strings are multiplied with input raw data block to get extracted bits. Binary multiplication between bits is realized by the `AND' operation, and binary addition between bits is realized by the `XOR' operation. The output bits obtained from each block are concatenated consecutively to produce the batch output bit-string. The above process is repeated until all the batches have been successfully extracted. The final random number output buffer is obtained as a long concatenated bit-string of binary bits. 

Both of the above-mentioned sub-sections are major processes that are performed during the post-processing of raw data from a QRNG on an FPGA as shown in Fig. 3. The real-time data acquisition of a quantum entropy source with the FPGA to perform post-processing includes working with a high-speed analog-to-digital converter (ADC) interfacing it with a memory block of the FPGA then retrieving the required sample raw data. Following this, we perform the entire post-processing of the sample raw data on an FPGA. Post acquiring the random number output buffer, we can interface the data via any of the high-speed interconnects available on the FPGA, like USB 3.0, Gigabit Ethernet, and PCI Express, to interface it with any display peripheral for further use. The i/o interfacing operations with hardware are not performed during the execution and testing of the current work. Instead, we focus solely on the operations that utilize the FPGA to perform post-processing involving min-entropy evaluation, output length calculation, Toeplitz string generation, randomness extraction via the Toeplitz strong extractor, and finally obtaining the random number output buffer. 

\begin{figure*}[htbp]
\includegraphics[width=1.6\columnwidth]{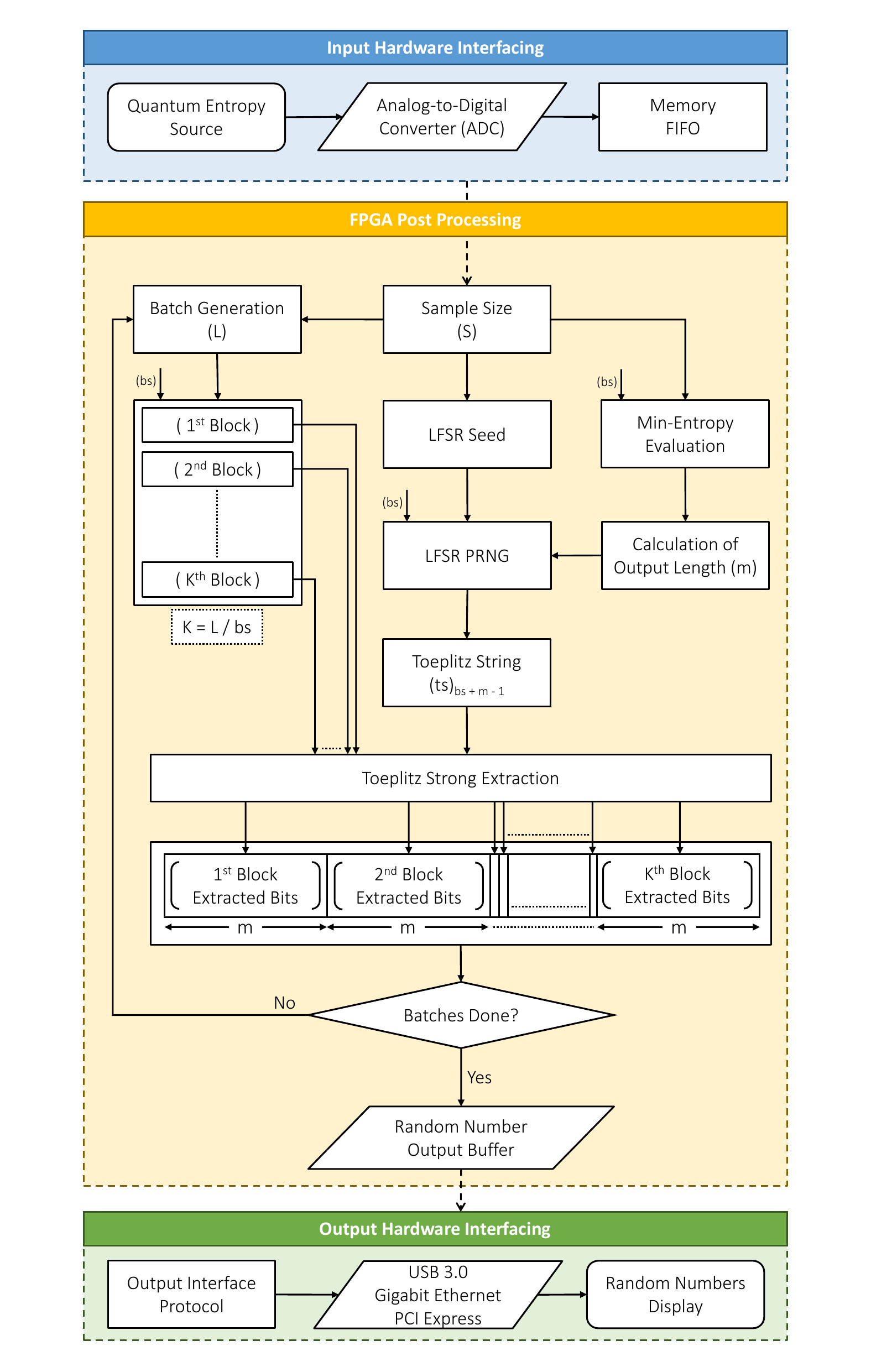}
\centering{\caption{Execution Flow of the Toeplitz Strong Extractor on Xilinx VC709 FPGA. (bs) denotes a standard input block size of $1 \times 10^{3}$ bits. The algorithm is performed for a sample size of $8 \times 10^{5}$ bits of raw QRNG data.}}
\label{fig3}
\end{figure*}

\section{Results and Discussion}

We use the system clock of the VC709 FPGA, which runs at $200$ MHz. The execution of the FPGA Post-Processing block shown in Fig. 3 has some one-time overheads, such as the min-entropy evaluation, and calculation of output length, followed by using the LFSR PRNG to generate the Toeplitz random string. These operations take a total of $100274$ clock cycles and pose a one-time overhead of $502 \mu s$.   

The number of clock cycles it takes to perform the Toeplitz Strong Extractor process to perform randomness extraction for various extraction ratios is shown in Table~\ref{table1}. The randomness extraction ratio $(ER)$ is the ratio of the output bits length $(m)$ to the input bits length $(bs)$.

\begin{table}[htbp]
\centering
\caption{Number of clock cycles for different randomness extraction ratios on VC709.}
\label{table1}
\begin{tabular}{|c|c|}
\hline
\textbf{Extraction Ratio} & \textbf{Clock Cycles} \\
\hline
0.3 & 6021 \\
0.5 & 10021 \\
0.6 & 12021 \\
0.8 & 16021 \\
\hline
\end{tabular}
\end{table}

\begin{figure}[htbp]
\centerline{\includegraphics[width=3.375in]{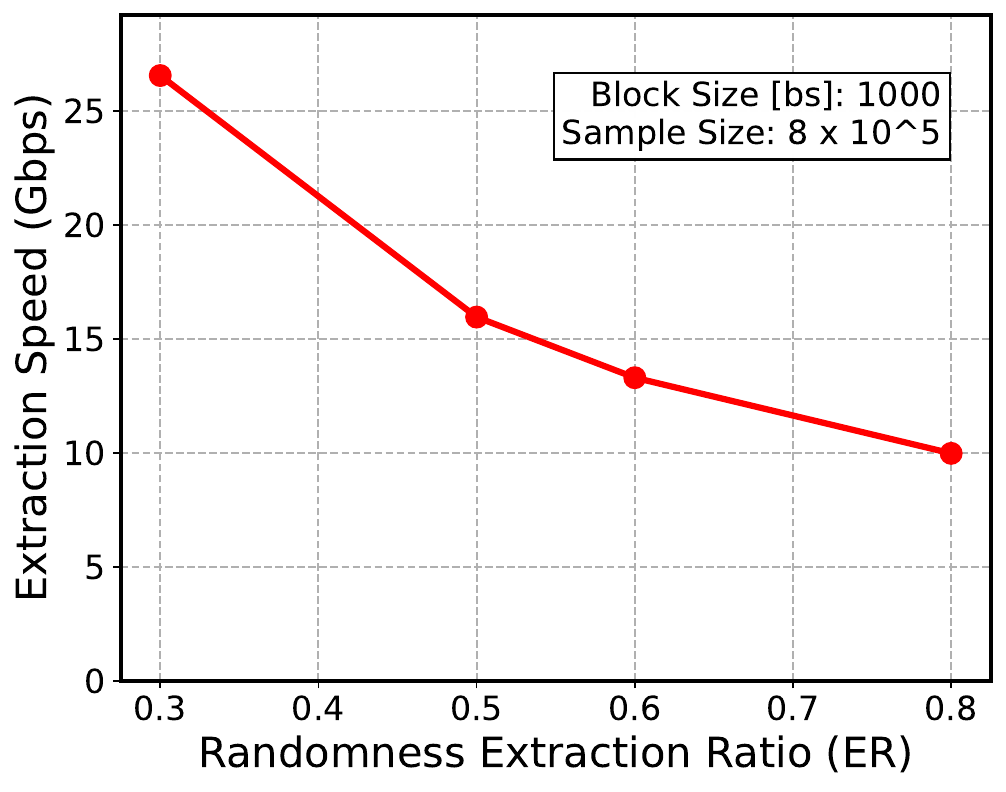}}
\caption{Extraction Ratio vs Randomness Extraction Speed in Gbps on Xilinx VC709.}
\label{fig4}
\end{figure}

Based on the number of clock cycles and the frequency of the clock, we calculate the time taken to perform the operation and use it to obtain the randomness extraction speed. We observe a steady decrease in the speed as the randomness extraction ratio increases, as shown in Fig.~\ref{fig4}. We achieved a maximum randomness extraction speed of $26.57$ Gbps at an ER of $0.3$ and a speed of $9.99$ Gbps at an ER of $0.8$.  

For the data considered as the sample size of $8 \times 10^{5}$ bits from the phase noise-based QRNG, the min-entropy evaluated is $2.6$ bits per $8$ bits which results in an output bit length $(m)$ of $300$ for the block size of $1000$ calculated based on the leftover hash lemma \cite{ma_2013_postprocessing, kollmitzer_2020_quantum}. The security parameter $(\varepsilon)$ considered for the calculation of `m' is $2^{-12.5}$. A randomness extraction speed of $26.57$ Gbps was achieved by applying the Toeplitz Strong Extractor on the QRNG's raw data. 

\begin{figure}[htbp]
\centerline{\includegraphics[width=3.375in]{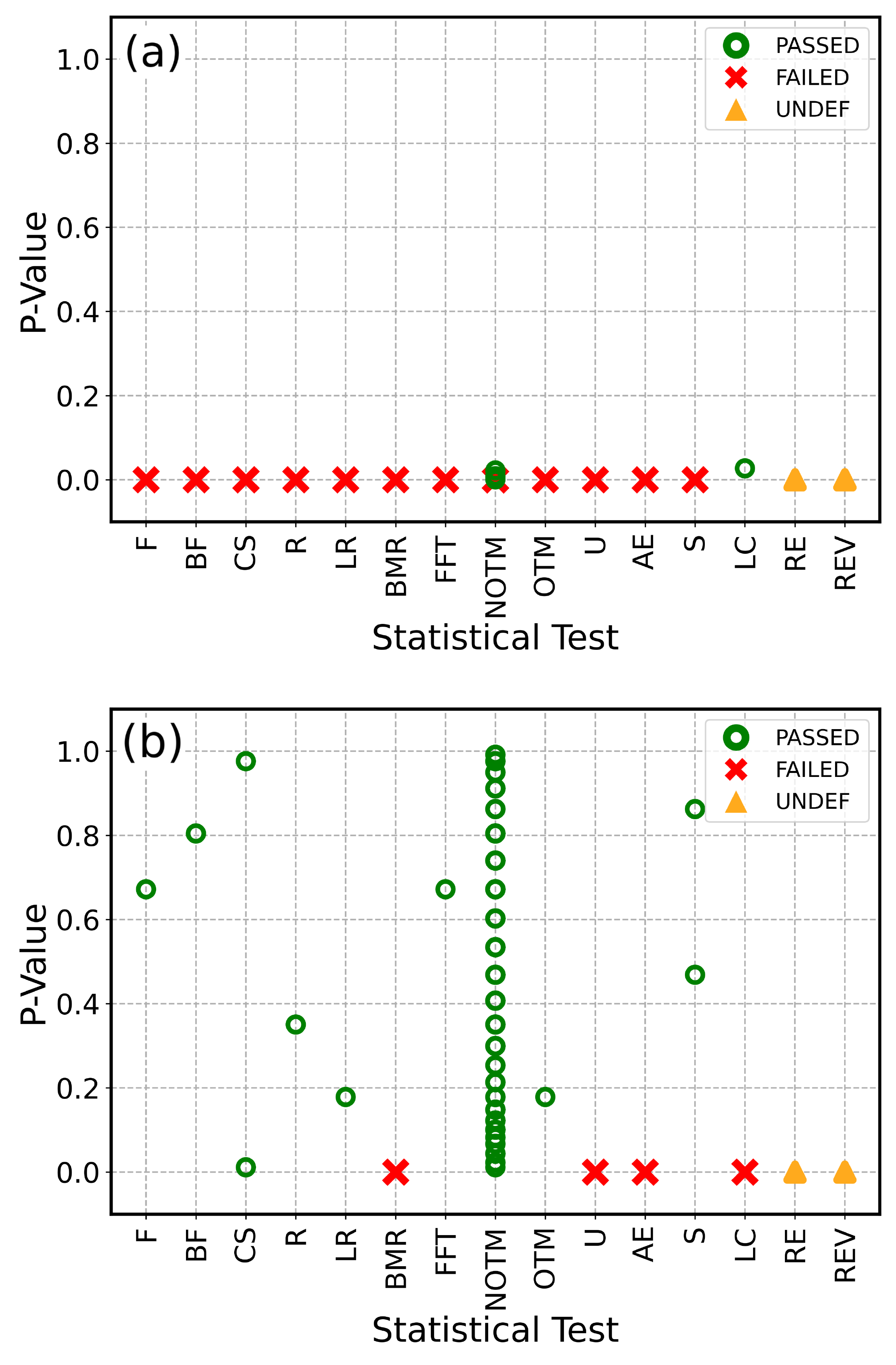}}
\caption{NIST STS 2.1.2 Randomness Tests Suite applied on (a) Raw Data of QRNG (b) Extracted Data of QRNG.}
\label{fig5}
\end{figure}

The QRNG's raw data and the output data from the extractor are then passed through the NIST STS 2.1.2 statistical randomness tests suite for verification of the post-processing. 800 Kbits of raw data were divided into 100 bit-strings of 8000 bits each, while 240 Kbits of extracted data were divided into 30 bit-strings of 8000 bits each and passed through the test suite. The results of the testing can be found in Fig.~\ref{fig5}. The lack of sufficient data led to tests such as the `Random Excursions (RE)' and the `Random Excursions Variant (REV)' tests returning undefined results. The post-extraction test failures may be attributed to similar limitations detailed in the NIST SP 800-22 official documentation \cite{nist_2010} as shown in Fig.~\ref{fig5}(b).

\section{Conclusion}

 The present work demonstrates an efficient and high-speed implementation of the Toeplitz Strong Extractor on an FPGA, achieving a new benchmark of $26.57$ Gbps with a randomness extraction ratio of $0.3$. A higher throughput implementation with a better randomness extraction ratio of $0.6$ gives us a speed of $13.3$ Gbps. A detailed execution flow of the process on an FPGA is presented, followed by performing the post-processing on raw data from an in-house QRNG and validating the extracted data output using the NIST STS 2.1.2 standardized statistical randomness tests suite. 

 Further studies may be performed by moving to the FFT-based Teoplitz Strong Extractor implementation, providing a better time and space complexity compared to the Matrix Multiplication-based Teoplitz Strong Extractor. Likewise, real-time post-processing speeds may be benchmarked across various output interfaces such as USB3.0, PCIe, and Gigabit Ethernet, enabling application-specific use-cases of the TSE on FPGA. Further increase in the extraction speeds may also be achieved by the implementation of a multi-channel FPGA extraction approach as shown in \cite{tanizawa_2024_realtime, guo_2024_parallel}.



\end{document}